*Subject Section*

# Shu: Visualization of high dimensional biological pathways

Jorge Carrasco Muriel[1,*], Nicholas Cowie[1], Marjan Mansouvar[2], Teddy Groves[1] and Lars Keld Nielsen[3]

[1]The Novo Nordisk Foundation Center for Biosustainability, Technical University of Denmark, DK-2800 Kgs. Lyngby, Denmark, [2]Department of Biotechnology and Biomedicine, Technical University of Denmark, DK-2800 Kgs. Lyngby, Denmark, [3]Australian Institute for Bioengineering and Nanotechnology (AIBN), The University of Queensland, St Lucia, QLD 4067, Australia.

*To whom correspondence should be addressed.

## Abstract

**Summary:** Shu is a visualization tool that integrates diverse data types into a metabolic map, with a focus on supporting multiple conditions and visualizing distributions. The goal is to provide a unified platform for handling the growing volume of multi-omics data, leveraging the metabolic maps developed by the metabolic modeling community. Additionally, shu offers a streamlined python API, based on the Grammar of Graphics, for easy integration with data pipelines.
**Availability and implementation:** Freely available at https://github.com/biosustain/shu under MIT/Apache 2.0 license. Binaries are available in the release page of the repository and the web app is deployed at https://biosustain.github.io/shu.
**Contact:** jcamu@biosustain.dtu.dk
**Supplementary information:** Supplementary data are available at *Bioinformatics* online.

## 1 Introduction

Systems Biology produces large amounts of data of varying nature across different experimental conditions (Cvijovic *et al.*, 2014). The Multi-omics paradigm starts from the premise that information pertaining to different biological layers - from DNA to metabolite abundances, fluxes and more - can link together synergistically, so that a linked dataset can generate knowledge beyond that which could be obtained from each individual data type. However, it remains a challenge to conceive and implement mathematical models that can accommodate it and then to devise informative visual representations of the complex interactions of the data (Agamah *et al.*, 2022).

In the field of genome-scale metabolic modeling, the metabolic maps provided by Escher (King *et al.*, 2015) have been a highly successful representation to depict the results of metabolic engineering (Masson *et al.*, 2023). Moreover, other software tools have built on top of or followed escher with other metabolic applications like Caffeine (*Caffeine*, no date), CNApy (Thiele *et al.*, 2022) or Omix (Droste, Nöh and Wiechert, 2013).

We aimed to address two outstanding problems with current metabolic network visualization tools. First, the existence of distributions of data in opposition to point estimates for a given biological entity. Existing metabolic maps rely on continuous color scales or sizes which cannot represent a distribution of data. This is a key challenge when dealing with high-throughput technologies and especially with the output of computational biology models that provide a distribution as a result (Medlock, Moutinho and Papin, 2020; Matos *et al.*, 2022). Second, the importance of the comparison between different biological conditions for the same experimental setup (knock-outs, changes in media, etc.). For example, differential omics experiments are very common in biology (Monti *et al.*, 2019) and their insights are naturally brought to light only when two or more conditions are considered.

In this application note we present shu: a software application that can be used both natively and in a web app to visualize different kinds of information on top of metabolic maps, with emphasis on the representation of distributions and different conditions. Additionally, we developed a Python package to enable programmatic access to shu.

## 2 The software

Shu is written in the Rust Programming Language (Matsakis and Klock, 2014) on top of the open-source bevy Entity-Component-System framework (Bevy Contributors, 2022).

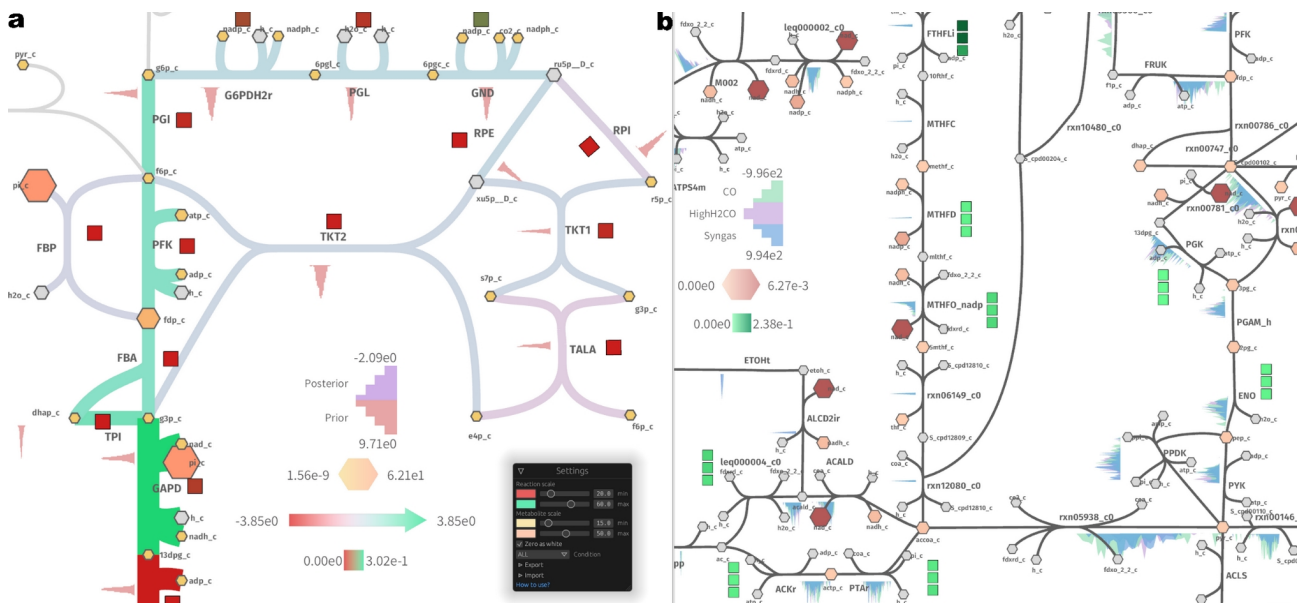

**Fig. 1. a.** *A posteriori* concentrations as size and color of metabolites, enzyme concentrations as box color and flux as color and size of arrows in absolute scale. Priors for log-k$_{cat}$ values are shown as red histograms. **b.** Metabolomics, proteomics and flux sampling plotted as the size and color of the metabolites; the color of the boxes; and the density curves respectively. Box conditions top to bottom: CO, H2-High CO and Syngas.

Shu requires two inputs: the metabolic map and the biological data. The metabolic map is compatible with Escher, thus making existing Escher maps made by the metabolic engineering community, covering many organisms and metabolic pathways, readily available for visualization using shu.

The biological data is mapped to the identifiers of metabolites, reactions, and conditions, and is represented by various geometric entities on the map. These include the color and size of arrows and nodes, histograms and kernel density estimates, and boxes at each side of the arrow. This allows for plotting 8 different data variables across conditions on the same map.

Histograms and density estimates can be plotted alongside arrows and nodes. For arrows, three positions are allowed to plot three different variables: the left side, the right side and the hovering window. The axis scale is consistent across all plots, allowing for easy comparison of the same variable across different reactions. Since the placement heuristics might fail for complex reactions, the axis can be moved, rotated and scaled freely using the mouse. The positions of the axis can be saved to a new map via the Graphical User Interface.

Boxes in shu refer to colored squares that depict a continuous color scale associated with a reaction and possibly a condition. The other geometrical shapes associated with color and size in arrows and nodes follow the same representation as in escher and other metabolic maps.

Shu allows for the integration of data with or without explicitly labeled experimental conditions. One or all conditions can be selected to be displayed. This is particularly useful for histograms, boxes, and density functions in publication-ready figures. Since arrow colors can only display one condition at a time, reaction data in different conditions can be mapped to boxes instead, sorted vertically to display all point estimates across all conditions. A legend is automatically generated only for the geometrical entities associated with data.

Given the possible complexity of the data that shu visualizes, a Python package called ggshu is made available to enable users to render this data programmatically.

Documentation for the usage of shu and ggshu is available at https://biosustain.github.io/shu/docs. The rust API is documented at https://docs.rs/shu. Binaries for Linux, Mac and Windows are distributed for every version and a Web Assembly application was compiled and deployed to be used directly in the browser at https://biosustain.github.io/shu.

As a final remark, it is important to note that we emphasize maps of metabolism because of the prior success in this area coupled with the easy availability of metabolic maps. However, shu is not tied to metabolism and could potentially be used for other kinds of networks, which are a pervasive representation in biology and beyond.

## 3 Case studies

Two case studies are presented to highlight the usefulness of shu for visualizing the output of computational models and experimental multi-omics data. One jupyter notebook for each case study is available in the supplementary material for the data integration with and without the python package. Data, models and maps are available at https://github.com/biosustain/shu_case_studies.

### 3.1 Visualization of *a posteriori* distributions from a bayesian kinetic model across n-conditions

The computational model is Maud (Maud contributors, 2022), a Bayesian kinetic model that outputs posterior distributions for kinetic and omics variables. A Maud kinetic model of 23 reactions, 27 metabolites and 24 enzymes of the central metabolism of *Escherichia coli* was used to generate *a posteriori* distributions of the kinetic parameters for simulated experimental data. In particular, k$_{cat}$ constants, enzyme concentrations, fluxes and metabolite concentrations were selected for visualization. Figure 1a reflects how these values were mapped, showing the "Settings" window for reference.

### 3.2 Proteomics, metabolomics and computational fluxes of *Clostridium autoethanogenum*

The experimental data is metabolomics, proteomics and computational fluxes for *Clostridium autoethanogenum* (de Souza Pinto Lemgruber *et al.*, 2018; Valgepea *et al.*, 2018, 2022). The proteomics and metabolomics datasets were collected for three conditions: "syngas", "high-H2 CO" and "CO" in the high biomass setup (1.4 gDCW/L). The genome-scale metabolic model of *C. autoethanogenum* iCLAU786 (Valgepea *et al.*, 2017) was used to generate flux distributions for the three different conditions by constraining the exchange fluxes of the byproducts and sampling distributions using optGp (Megchelenbrink, Huynen and Marchiori, 2014) method in COBRApy (Ebrahim *et al.*, 2013). These were mapped to densities in the map.

The result is shown in [Figure 1B]. The enzyme concentrations were depicted in the left boxes while the metabolite concentrations were mapped as the color and size of the nodes. The map makes it easy to reason globally about which pathways are more constrained; which proteins vary across conditions; or how the NADH concentration compares with NAD or NADPH. At the same time, individual reactions can be inspected to make local observations about, for instance, the irreversibility of a reaction, like ENO or ACKr.

*shu*

## Funding

This work has been supported by the Novo Nordisk Foundation (NNF Grant numbers: NNF20CC0035580 and NNF14OC0009473).

*Conflict of Interest:* none declared.